\documentclass[english,final,3p,times,twocolumn]{elsarticle}
\usepackage[T1]{fontenc}
\usepackage[latin9]{inputenc}
\usepackage{amsmath}
\usepackage{graphicx}
\usepackage{amssymb}

\makeatletter
\usepackage{amsbsy}

\usepackage{babel}
\usepackage{lineno}

\journal{Solid State Communications}

\makeatother

\usepackage{babel}

\begin{document}

\title{Magnetocaloritronic nanomachines}

\author{Alexey A. Kovalev}

\author{Yaroslav Tserkovnyak}

\address{Department of Physics and Astronomy, University of California, Los
Angeles, California 90095, USA}
\begin{abstract}
We introduce and study a magnetocaloritronic circuit element based
on a domain wall that can move under applied voltage, magnetic field
and temperature gradient. We draw analogies between the Carnot machines
and possible devices employing such circuit element. We further point
out the parallels between the operational principles of thermoelectric
and magnetocaloritronic cooling and power generation and also introduce
a magnetocaloritronic figure of merit. Even though the magnetocaloritronic
figure of merit turns out to be very small for transition-metal based
magnets, we speculate that larger numbers may be expected in ferromagnetic
insulators. \end{abstract}
\maketitle

\section{Introduction}

There have been numerous realizations of Carnot machines since they
were first envisaged by Nicolas L{é}onard Sadi Carnot, in both direct
(\textit{i.e.}, engine) and reverse (\textit{i.e.}, refrigerator or
heat pump) modes of operation. While the traditional mechanical Carnot
machines are based on the alternating adiabatic and isothermal processes
controlled by the conjugate pair of variables ($P$,$V$), same thermodynamic
principles can be put to work in the realizations of magnetic machines
relying on the magnetocaloric effect \citep{Pecharsky:oct1999}. These
latter machines are operating in the space of the conjugate variables
($H$,$M$). At nanoscale, however, one might need to rely on different
principles and thermoelectric cooling and power generation appear
to be very promising \citep{DiSalvo:jul1999,Ohta:feb2007}. In particular,
large values of the thermoelectric figure of merit have recently been
suggested for molecular junctions \citep{Murphy:oct2008}. In this
paper, we envision an interplay of the magnetocaloric and thermoelectric
functionalities.

A growing interest in spin caloritronics that comprises the spin related
phenomena with thermoelectric effects has been spurred recently by
many promising applications \citep{Sales:feb2002,Hochbaum:jan2008,Boukai:jan2008}.
Thermoelectric spin transfer relates the heat current to magnetization
dynamics \citep{Hatami:aug2007,Hatami:may2009,Kovalev:sep2009} while
opposite effect of heat currents resulting from magnetization dynamics
should also occur \citep{Kovalev:sep2009,Bauer:unpublished}. The
spin-transfer torque \citep{Berger:Oct1996,Slonczewski:1996} in spin
valves and domain walls \citep{Yamaguchi:Feb2004,Hayashi:may2006,Hayashi:jan2007}
has been well understood for transition-metal based magnets \citep{Tatara:2008,Ralph:2008,Tserkovnyak:apr2008a}
which already led to many applications \citep{Parkin:2008}. The reciprocal
effect to the spin-transfer torque results in electromotive forces
induced by the magnetization dynamics \citep{Barnes:jun2007,Tserkovnyak:apr2008,Duine:2008,Saslow:2007}.
All these pave the way for novel devices that can output as well as
be controlled by temperature gradients, electric currents, and magnetic
fields. These machines can have similar functionalities as Carnot
machines and work at nanoscale. 

In this paper, we introduce and describe a magnetocaloritronic circuit
element utilizing magnetic domain wall motion. Further, we use this
element to demonstrate the principle of magnetocaloritronic cooling
and power generation. We also draw parallels between the operational
principles of thermoelectric and magnetocaloritronic cooling and power
generation. This program naturally leads us to the introduction of
the magnetocaloritronic figure of merit $TZ_{\mbox{mc}}$ which encodes
information about the maximum efficiency of such devices. Our estimates
indicate a very small figure of merit for typical transition-metal
based magnets; however, we speculate that one can achieve better efficiencies
using the ferromagnetic insulators in which the heat transferred by
spin waves will better couple to the texture dynamics in the absence
of dissipation related to the electron-hole continuum.

\section{Magnetocaloritronic circuit element}

\begin{figure}
\includegraphics[clip,width=1\columnwidth]{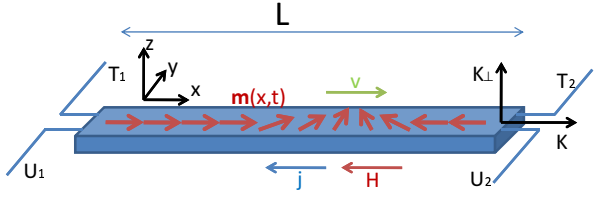}

\caption{(Color online) A domain-wall based circuit element can be controlled
by applying voltage $\Delta U$, magnetic field $H$ and temperature
gradient $\Delta T$. Here, we consider transverse head-to-head N{é}el
domain wall parallel to the $y$ axis in the easy $xy$ plane. The
constants $K$ and $K_{\perp}$ describe the easy axis and easy plane
anisotropy.}

\label{fig1} 
\end{figure}

In order to explore new functionalities, we introduce a domain-wall
based circuit element (see Fig. \ref{fig1}) that combines the capabilities
of a thermoelectric contact, heat pump and generator of electromagnetic
field. The functionalities of this circuit element can be controlled
directly by applying magnetic field which leads to domain wall motion
along the magnetic field in the direction of the lower free energy.
Alternatively, one can control the domain wall by applying voltage
and temperature gradients which also couple to the domain wall motion
through the viscous interaction of the charge and energy currents
with the magnetization dynamics. The velocity of the domain wall then
becomes:\begin{equation}
\upsilon=\dfrac{\gamma HW-p\beta j/s-p^{'}\beta^{'}j_{q}/s}{\alpha}\:,\label{DW-velocity}\end{equation}
where the domain wall acquires some velocity in response to the magnetic
field $H$, the charge current $j$ and the energy current $j_{q}$.
Equation (\ref{DW-velocity}) is derived below in this section from
the Landau-Lifshitz-Gilbert (LLG) equation for the Walker ansatz along
with the introduction of the coupling constants $p$, $p^{'}$, $\beta$
and $\beta^{'}$. The other parameters are the Gilbert damping constant
$\alpha$, the domain wall width $W$, the spin density $s$ where
$s\mathbf{m}=\mathbf{M}/\gamma$, with $M$ being the magnetization
density, $\mathbf{m}$ the unit vector along the spin density and
$\gamma$ the gyromagnetic ratio ($\gamma<0$ for electrons). 

By writing the equation for the entropy production in the form analogous
to the microscopic form in Ref. \citep{Kovalev:sep2009}: \begin{equation}
\dot{\mathbb{S}}=\dfrac{L}{T}\left[-j_{q}\dfrac{T_{2}-T_{1}}{TL}-j\dfrac{\mu_{2}-\mu_{1}}{L}-\dot{X}\dfrac{2MH}{L}\right]\,,\label{Entropy_Rate_1}\end{equation}
and identifying the thermodynamics variables $j_{q}$, $j$ and $\dot{X}$,
we can phenomenologically generalize Eq. (\ref{DW-velocity}) to systems
with the most general macroscopic equations describing domain wall
dynamics \citep{Bauer:unpublished}:\begin{equation}
\begin{array}{c}
\dot{X}=-\mathcal{O}_{X}\dfrac{2MH}{L}-\mathcal{O}_{X\mu}j-\mathcal{O}_{XT}j_{q}\:,\\
\dfrac{T_{2}-T_{1}}{TL}=-\mathcal{O}_{T}j_{q}+\mathcal{O}_{XT}\dfrac{2MH}{L}\:,\\
\dfrac{\mu_{2}-\mu_{1}}{L}=-\mathcal{O}_{\mu}j+\mathcal{O}_{X\mu}\dfrac{2MH}{L}\:,\end{array}\label{Onsager}\end{equation}
where $X$ is the position of the domain wall, $T=(T_{1}+T_{2})/2$
and the last $2$ equations have been inverted with respect to $j$
and $j_{q}$ for convenience of the following derivations. We introduced
$5$ kinetic coefficients $\mathcal{O}_{X}$, $\mathcal{O}_{T}$,
$\mathcal{O}_{\mu}$, $\mathcal{O}_{XT}$ and $\mathcal{O}_{X\mu}$
in accordance with the Onsager principle.

We will now turn to the microscopic derivation of Eqs. (\ref{Onsager})
for the case of transition-metal based magnets keeping in mind that,
in principle, these equations can be valid for other systems as well
(\textit{e.g.} for magnetic semiconductors in Ref. \citep{Hals:jun2009}).
The dynamics of the circuit element in Fig. \ref{fig1} can be conveniently
described by the following generalization to the Landau-Lifshitz-Gilbert
(LLG) equation \citep{Kovalev:sep2009}:\begin{equation}
\partial_{x}\mu=-gj+\xi j_{q}+p\left(\mathbf{m}\times\partial_{x}\mathbf{m}+\beta\partial_{x}\mathbf{m}\right)\cdot\dot{\mathbf{m}}\,,\label{ohmD1}\end{equation}
\begin{equation}
\partial_{x}T/T=\xi j-\zeta j_{q}+p^{'}\left(\mathbf{m}\times\partial_{x}\mathbf{m}+\beta^{'}\partial_{x}\mathbf{m}\right)\cdot\dot{\mathbf{m}}\,,\label{HeatCurr1}\end{equation}
\begin{equation}
\begin{array}{c}
s(1+\alpha\mathbf{m}\times)\dot{\mathbf{m}}+\mathbf{m}\times\mathbf{H}=p\left(\partial_{x}\mathbf{m}+\beta\mathbf{m}\times\partial_{x}\mathbf{m}\right)j\\
\hspace{1.5cm}+p^{'}\left(\partial_{x}\mathbf{m}+\beta^{'}\mathbf{m}\times\partial_{x}\mathbf{m}\right)j_{q}\,,\end{array}\label{LLGD-heat1}\end{equation}
where $j$ is the charge current and $j_{q}=j_{U}-\mu j$ is the energy
current offset by the energy corresponding to the chemical potential
$\mu$ with $j_{U}$ being the ordinary energy current. The kinetic
coefficients $g$, $\xi$, $\zeta$ and $\alpha$ can in general also
depend on temperature and texture in isotropic materials, for the
latter, to the leading order, as $g=g_{0}+\eta_{g}(\partial_{x}\mathbf{m})^{2}$,
$\xi=\xi_{0}+\eta_{\xi}(\partial_{x}\mathbf{m})^{2}$, etc.. The coefficients
$p$ and $p'$ describe the so called nondissipative \citep{Tserkovnyak:jan2009,Kovalev:sep2009}
coupling of the magnetization dynamics to the charge and energy currents.
The corresponding viscous corrections due to electron spin's mistracking
of the magnetic texture are described by the coefficients $\beta$
and $\beta^{'}$ \citep{Tserkovnyak:jan2009,Kovalev:sep2009}. The
coefficients $g$, $\xi$ and $\zeta$ can be related to the thermal
conductivity $\kappa=1/\zeta T$, the Peltier coefficient $\Pi=\xi/\zeta$
and the conductivity $1/\sigma=g-\Pi^{2}/\kappa T$. In general, $\mathbf{H}$
is different from the usual {}``effective field'' corresponding
to the variation of the Landau free-energy functional $F[\mathbf{m},\mu,T]$
with respect to $\mathbf{m}$ at a fixed $\mu$ and $T$, and can
be expanded phenomenologically in terms of small $\partial_{x}T$
and $\partial_{x}\mu$ \citep{Kovalev:sep2009}. In order to avoid
unnecessary complications, we assume here that even in an out-of-equilibrium
situation, when $\partial_{x}T\neq0$ and $\partial_{x}\mu\neq0$,
$\mathbf{H}$ depends only on the instantaneous texture $\mathbf{m}(x)$
so that $\mathbf{H}\equiv\partial_{\mathbf{m}}F$. The texture corrections
in Eqs. (\ref{ohmD1}) and (\ref{HeatCurr1}) modify the energy and
charge flows and can be relatively large in some cases \citep{Kovalev:sep2009}
leading to texture corrections of the Gilbert damping \citep{Tserkovnyak:jan2009}
in the LLG Eq. (\ref{LLGD-heat1}). However, for sufficiently smooth
domain walls, these corrections to the Gilbert damping are small and
will be disregarded hereafter. The parameters $p$ and $p^{'}$ can
be approximated in the strong exchange limit as \citep{Kovalev:sep2009}:\begin{equation}
p^{'}=\dfrac{\hbar}{2e}\wp_{S}\Pi_{0}\dfrac{\sigma_{0}(1-\wp^{2})}{T\kappa_{0}}\,,\; p=\dfrac{\wp\hbar}{2e}-p^{'}\Pi_{0}\,,\label{ThermalPolarization}\end{equation}
where $\kappa_{0}$, $\sigma_{0}$ and $\Pi_{0}$ are the thermal
and ordinary conductivities and the Peltier coefficient defined in
the absence of textures, respectively. The polarizations are defined
as $\wp=(\sigma_{0}^{\uparrow}-\sigma_{0}^{\downarrow})/\sigma_{0}$
and $\wp_{S}=(\Pi_{0}^{\uparrow}-\Pi_{0}^{\downarrow})/(\Pi_{0}^{\uparrow}+\Pi_{0}^{\downarrow})=e\Pi_{s}/2\Pi_{0}$
where $\Pi_{s}=(\Pi_{0}^{\uparrow}-\Pi_{0}^{\downarrow})/e$ is the
spin Peltier coefficient, $\sigma_{0}=\sigma_{0}^{\uparrow}+\sigma_{0}^{\downarrow}$,
and $-e$ is the charge of particles ($e>0$ for electrons).

We will describe the domain wall in Fig. \ref{fig1} by the Walker
ansatz valid for weak field and current biases \citep{Schryer:1974,Li:Jul2004,Tserkovnyak:apr2008a}:\begin{equation}
\varphi(\mathbf{r},t)\equiv\Phi(t),\quad\ln\tan\dfrac{\theta(\mathbf{r},t)}{2}\equiv\dfrac{x-X(t)}{W(t)}\:,\label{Walker}\end{equation}
where the position-dependent spherical angles $\varphi$ and $\theta$
parametrize the magnetic configuration as $\mathbf{m}=(\cos\theta,\sin\theta\cos\varphi,\sin\theta\sin\varphi)$,
$X(t)$ parametrizes the net displacement of the wall along the $x$
axis, and we assume that the driving forces ($H$, $j$ and $j_{q}$)
are not too strong so that the wall preserves its shape and only its
width $W(t)$ and out-of-plane tilt angle $\Phi(t)$ undergo small
changes. By substituting the ansatz (\ref{Walker}) in Eq. (\ref{LLGD-heat1})
with the effective field given by\[
\mathbf{H}=(H+km_{x})\mathbf{x}-K_{\perp}m_{z}\mathbf{z}+A\boldsymbol{\bigtriangledown}^{2}\mathbf{m},\]
we obtain: \begin{equation}
\begin{array}{c}
\dot{\Phi}+\dfrac{\alpha\dot{X}}{W}=\gamma H-\dfrac{p\beta j}{sW}-\dfrac{p^{'}\beta^{'}j_{q}}{sW}\:,\\
\dfrac{\dot{X}}{W}-\alpha\dot{\Phi}=\dfrac{\gamma K_{\perp}\sin2\Phi}{2}-\dfrac{pj}{sW}-\dfrac{p^{'}j_{q}}{sW}\:,\\
W=\sqrt{\dfrac{A}{K+K_{\perp}\sin^{2}\Phi}}\:,\end{array}\label{Walker1}\end{equation}
where $A$ is the stiffness constant and $K$ and $K_{\perp}$ describe
the easy axis and easy plane anisotropies, respectively. The steady
state solution of Eq. (\ref{Walker1}) below the Walker breakdown
with $\Phi(t)=\mbox{const}$ and $X=\upsilon t$ leads to the result
in Eq. (\ref{DW-velocity}). By comparing Eq. (\ref{Onsager}) with
Eqs. (\ref{ohmD1}), (\ref{HeatCurr1}) and (\ref{LLGD-heat1}), we
can also write the following expressions for the kinetic coefficients:\[
\begin{array}{c}
\mathcal{O}_{X}=\dfrac{LW}{2\alpha s}\:,\:\mathcal{O}_{T}+\dfrac{\mathcal{O}_{XT}^{2}}{\mathcal{O}_{X}}=\bar{\zeta}\:,\:\mathcal{O}_{\mu}+\dfrac{\mathcal{O}_{X\mu}^{2}}{\mathcal{O}_{X}}=\bar{g}\:,\\
\mathcal{O}_{X\mu}=\dfrac{p\beta}{\alpha s}\:,\:\mathcal{O}_{XT}=\dfrac{p^{'}\beta^{'}}{\alpha s}\:,\end{array}\]
where the coefficients $\bar{g}$ and $\bar{\zeta}$ correspond to
$g$ and $\zeta$ averaged over the wire.

\section{Magnetocaloritronic cooling}

The circuit element in Fig. \ref{fig1} can in principle be used as
a working body of a magnetic refrigerator exploiting the magnetocaloric
effect \citep{Pecharsky:oct1999}. Realization of such a refrigerator,
however, necessitates some sort of switch connecting and disconnecting
the circuit element to reservoirs. 

\begin{figure}
\includegraphics[clip,width=1\columnwidth]{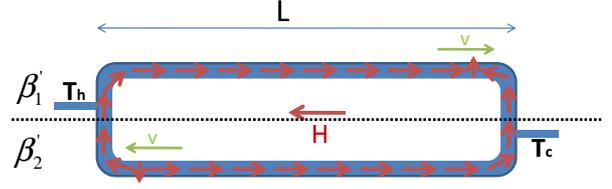}

\caption{(Color online) Magnetocaloritronic cooling can be realized by moving
domain walls between two regions with different viscous coupling $\beta^{'}$.
The cyclic motion of the domain walls is maintained by the rotating
clockwise magnetic field (along a horizontally elongated elliptical
trajectory) in sync with the domain walls steadily circulating clockwise;
this process is analogous to a pump which produces a dc heat flow
between the cold ($T_{c}$) and hot ($T_{h}$) junctions. The analogy
to the thermoelectric cooling based on p- and n-type couples \citep{DiSalvo:jul1999}
can be best seen when $p_{1}^{'}\beta_{1}^{'}=-p_{2}^{'}\beta_{2}^{'}$.}

\label{fig2} 
\end{figure}

At nanoscale, such switches can be hard to realize and in this work,
we will explore a different route that has more parallels to thermoelectric
cooling \citep{DiSalvo:jul1999} in which the cooling effect appears
at the junction formed by two different conducting materials. In our
case (see Fig. \ref{fig2}), the charge carriers are replaced by domain
walls sliding along the wire due to rotating magnetic field which
ensures cyclic motion of domain walls. As it can be obtained from
Eqs. (\ref{ohmD1}) and (\ref{HeatCurr1}), the heat transfer originates
from differences in the viscous coupling $\beta^{'}$ in the upper
and lower parts of the circuit in Fig. \ref{fig2} which can be a
result of different amounts of magnetic impurities in the corresponding
parts. 

It is customary to describe the efficiency of thermoelectric circuits
by a material figure of merit $Z=\Pi^{2}\sigma/T^{2}\kappa$ corresponding
to the maximum achievable temperature difference $\sim ZT^{2}$. In
most circumstances, a dimensionless figure, $ZT$, is quoted and it
corresponds to the relative temperature difference. Let us formulate
an analog of figure of merit for magnetocaloritronic device. We consider
the case of relatively small temperature gradients $T_{h}-T_{c}\ll T_{h}$.
In order to maintain the domain wall motion, we have to perform the
work $2MHL\mathcal{A}$ per one pass and all this work is eventually
dissipated, where $H$ is the applied field along the magnetic wires
and $\mathcal{A}$ is the cross-section of the wire. From Eq. (\ref{Onsager}),
we find that in the absence of temperature bias, the domain wall induces
the heat current: \begin{equation}
j_{\mbox{tr}}=\dfrac{\mathcal{O}_{XT}}{\mathcal{O}_{T}}\dfrac{2MH}{L}\:.\label{WallTransfer}\end{equation}
 For simplicity, we suppose that the charge current can not flow in
the device in Fig. \ref{fig2} (\textit{e.g.} due to a breaking point
or upper and lower parts could be different p- and n-type semiconductors,
whose junctions block the current flow). Supposing that the dissipated
work is equally distributed between the reservoirs, the externally
induced heat flow from the cold/hot reservoir per one wire then becomes:\begin{equation}
\begin{array}{c}
j_{q}^{\mbox{cold}}=\dfrac{T_{h}-T_{c}}{TL\mathcal{O}_{T}}+\dfrac{\mathcal{O}_{XT}}{\mathcal{O}_{T}}\dfrac{2MH}{L}+HM\dot{X}\:,\\
j_{q}^{\mbox{hot}}=\dfrac{T_{h}-T_{c}}{TL\mathcal{O}_{T}}+\dfrac{\mathcal{O}_{XT}}{\mathcal{O}_{T}}\dfrac{2MH}{L}-HM\dot{X}\:.\end{array}\label{HotCold}\end{equation}
Because the domain wall cooling is proportional to the magnetic field
and dissipative heating is proportional to the magnetic field squared,
the maximum decrease in the junction temperature is obtained at an
optimal magnetic field. This situation is similar to thermoelectric
cooling in which the Peltier cooling is proportional to current and
Joule heating is proportional to the current squared, leading to existence
of an optimal current. From Eq. (\ref{HotCold}), the maximum temperature
difference that the domain wall motion can maintain becomes:\[
(T_{h}-T_{c})^{\mbox{max}}=\dfrac{T\mathcal{O}_{XT}^{2}}{2\mathcal{O}_{X}\mathcal{O}_{T}}\:.\]
We can finally introduce the figure of merit for the magnetocaloritronic
cooling:\begin{equation}
TZ_{\mbox{mc}}=\dfrac{\mathcal{O}_{XT}^{2}}{\mathcal{O}_{X}\mathcal{O}_{T}}=\dfrac{1}{\dfrac{\alpha sWL\bar{\zeta}}{2p^{'2}\beta^{'2}}-1}\:,\label{MaxEfficiency}\end{equation}
where the last part of the equation is written for a domain wall described
by microscopic Eqs. (\ref{ohmD1}), (\ref{HeatCurr1}) and (\ref{LLGD-heat1}).
Notice that $TZ_{\mbox{mc}}>0$ as it follows from the thermodynamic
inequalities $\mathcal{O}_{T}\geq0$ and $\mathcal{O}_{X}\geq0$ which
guarantee that the entropy production in Eqs. (\ref{Entropy_Rate_1})
is always positive. It is also worthwhile to note that the efficiency
$TZ_{\mbox{mc}}$ becomes infinite when $\bar{\zeta}=2p^{'2}\beta^{'2}/WL\alpha s$
which corresponds exactly to the lower bound of the thermal resistivity
\citep{Kovalev:sep2009}, $\zeta=\eta_{\zeta}(\partial_{x}\mathbf{m})^{2}$
with $\eta_{\zeta}=\beta'^{2}p'^{2}/\alpha s$, averaged over the
Walker ansatz. 

Above, we only considered dissipation in the lower wire in Fig. \ref{fig2}.
Consideration of the upper wire does not change our results when the
upper wire is mirror symmetric to the lower with $p_{1}^{'}\beta_{1}^{'}=-p_{2}^{'}\beta_{2}^{'}$
(\textit{e.g.} we do not see any principal contradictions in existence
of negative $\beta^{'}$). In a different scenario when $\beta_{1}^{'}=0$
and $\beta_{2}^{'}\neq0$, it is possible to minimize the effect of
dissipation in the upper part by keeping the upper wire disconnected
from the cold/hot junction most of the time apart from the moments
when the domain wall moves through. 

It is interesting to see that a system in Fig. \ref{fig2} made of
p- and n-type semiconductors for upper and lower parts will have opposite
$p\beta$ in those parts leading to a device that can generate electrical
power from a rotating magnetic field in accordance with Eqs. (\ref{Onsager}).

At last, we recover the well known expressions for the maximum coefficient
of performance ($COP$) \citep{Mahan:1997} written for the magnetocaloritronic
cooling and heating:\begin{equation}
COP_{\mbox{heat}}=\dfrac{j_{q}^{\mbox{hot}}}{2HM\dot{X}}=\dfrac{T_{h}}{T_{h}-T_{c}}\dfrac{\sqrt{1+TZ_{\mbox{mc}}}-T_{c}/T_{h}}{\sqrt{1+TZ_{\mbox{mc}}}+1}\:,\label{MaxHeat}\end{equation}
\begin{equation}
COP_{\mbox{cool}}=\dfrac{j_{q}^{\mbox{cold}}}{2HM\dot{X}}=\dfrac{T_{c}}{T_{h}-T_{c}}\dfrac{\sqrt{1+TZ_{\mbox{mc}}}-T_{h}/T_{c}}{\sqrt{1+TZ_{\mbox{mc}}}+1}\:,\label{MaxCool}\end{equation}
where we maximized the above equations with respect to $H$ at a fixed
temperature bias. The Carnot efficiency is recovered when $TZ_{\mbox{mc}}\rightarrow\infty$.

Using Eq. (\ref{ThermalPolarization}), we can express the magnetocaloritronic
figure of merit via the thermoelectric figure of merit: \[
TZ_{\mbox{mc}}=\dfrac{1}{\dfrac{\alpha sWL}{ZT\left[\beta^{'}\wp_{S}(1-\wp^{2})\hbar/2e\right]^{2}\sigma}-1}\:,\]
which gives a very small number $Z_{\mbox{mc}}\approx10^{-7}Z$ for
a typical domain wall in a Py wire at room temperature with $L=2W=100\:\mbox{nm}$.
Much more pronounced cooling effect could be seen in MnSi below $30$K
for which we obtain $Z_{\mbox{mc}}\approx10^{-3}Z$ (MnSi parameters
are taken to be the same with Ref. \citep{Kovalev:sep2009}). We speculate
that one can achieve better efficiencies using the ferromagnetic insulators
in which the heat transferred by spin waves will better couple to
the texture dynamics in the absence of dissipation related to the
electron-hole continuum. Large viscous $\beta$-like coupling with
domain wall motion have recently been predicted for dirty (Ga,Mn)As
\citep{Hals:jun2009} which can lead to larger $TZ_{\mbox{mc}}$ according
to Eq. (\ref{MaxEfficiency}). Note that the best materials available
today for devices that operate near room temperature have a $ZT$
of about $1\div2$ \citep{DiSalvo:jul1999,Harman:may2005}. 

\begin{figure}
\includegraphics[clip,width=1\columnwidth]{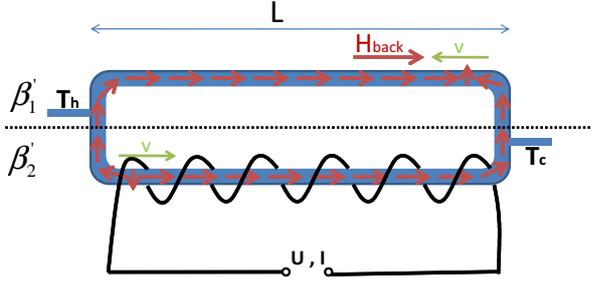}

\caption{(Color online) Magnetocaloritronic power generation can be realized
by maintaining cyclic motion of domain walls between two regions with
different viscous coupling $\beta^{'}$. The cyclic motion of the
domain walls can be maintained by the temperature gradient when $p^{'}\beta^{'}$
has opposite sign for the upper and lower parts. Alternatively, when
$\beta_{1}^{'}=0$, a small magnetic field $H_{\mbox{back}}$ can
return the domain walls to the hot junction. The power (useful work)
is extracted from a solenoid encircling one of the wires.}

\label{fig3} 
\end{figure}

\section{Magnetocaloritronic power generation}

Thermoelectric devices find numerous applications as voltage generators
\citep{DiSalvo:jul1999}. In this section, continuing the analogy
between the thermoelectric devices and the magnetocaloritronic devices,
we will show how power and useful work can be generated magnetocaloritronically.
The device depicted in Fig \ref{fig3} once again contains two regions
with different viscous coupling $\beta^{'}$. The temperature gradient
propels domain walls in the lower part with $\beta^{'}\neq0$ as it
follows from Eq. \ref{Onsager} while in the upper part with $\beta^{'}=0$
the domain walls are inert to the temperature gradient and move due
to the presence of a very small magnetic field $H_{\mbox{back}}$.
As an alternative, one could also consider the mirror symmetric case
with $p_{1}^{'}\beta_{1}^{'}=-p_{2}^{'}\beta_{2}^{'}$ and with an
additional solenoid encircling the upper wire.

Let us calculate the ratio of the useful power to the losses (efficiency)
for a magnetocaloritronic device. We again suppose that the charge
current can not flow in the device in Fig. \ref{fig3} (\textit{e.g.}
due to a breaking point or upper and lower parts could be different
p- and n-type semiconductors also blocking the current flow) and we
consider the case of relatively small temperature gradients $T_{h}-T_{c}\ll T_{h}$.
The dissipation due to domain wall motion $2MHL\mathcal{A}$ is again
evenly distributed along the wire and between reservoirs as we make
similar assumptions with the previous section. The losses then appear
due to the finite heat conductivity $j_{q}^{\mbox{cond}}=(T_{h}-T_{c})/TL\mathcal{O}_{T}$
and due to domain wall motion. The motion of domain wall through the
solenoid will lead to electromotive force in the solenoid $U=8\pi M\mathcal{A}N\dot{X}/L$
where $N$ is the number of loops in the solenoid and $\dot{X}$ is
the average velocity of the domain wall that can be found from Eq.
(\ref{Onsager}) with the magnetic field inside of the solenoid $H=4\pi IN/L$.
One can write the following expression for the useful power:\[
UI=2M\mathcal{A}\dot{X}H\:,\]
which, as in the previous section, suggests the existence of the optimal
$H$ for the maximum power outcome. We can finally recover the well
known expression for the maximum thermoelectric efficiency of power
generation \citep{Mahan:1997} written for the magnetocaloritronic
power efficiency. By maximizing the ratio of the power (work) $UI$
to the losses (heat absorbed at the hot end given by Eq. (\ref{HotCold}))
at a fixed temperature bias for the device in Fig. \ref{fig3}, we
obtain:\begin{equation}
\eta=\dfrac{UI}{j_{q}^{\mbox{hot}}\mathcal{A}}=\dfrac{T_{h}-T_{c}}{T_{h}}\dfrac{\sqrt{1+TZ_{\mbox{mc}}}-1}{\sqrt{1+TZ_{\mbox{mc}}}+T_{c}/T_{h}}\:.\label{MEfficiency}\end{equation}
As expected, the magnetocaloritronic figure of merit also contains
information about the efficiency of the magnetocaloritronic device
working as a power (useful work) generator. The Carnot efficiency
is once again recovered when $TZ_{\mbox{mc}}\rightarrow\infty$.

In the above estimates, we again considered dissipation only in the
lower wire in Fig. \ref{fig3}. Consideration of the upper wire does
not change our results when the upper wire contains an extra solenoid
and is mirror symmetric to the lower with $p_{1}^{'}\beta_{1}^{'}=-p_{2}^{'}\beta_{2}^{'}$.
In the scenario when $\beta_{1}^{'}=0$ and $\beta_{2}^{'}\neq0$,
a small magnetic field $H_{\mbox{back}}$ returns the domain walls
to the hot junction and we minimize the effect of dissipation in the
upper part by keeping the upper wire disconnected from the hot junction
most of the time apart from the moments when the domain wall moves
through.

\section{Conclusions}

We introduced and described the magnetocaloritronic circuit element
using recently formulated phenomenological theory of thermoelectric
spin transfer \citep{Kovalev:sep2009}. The velocity of domain wall
in such a circuit element in response to magnetic field, charge current
and energy flow is calculated. We also derive the most general form
of phenomenological equations describing dynamics of a domain wall
in response to the magnetic field, applied voltage and temperature
biases. We conclude that such hybrid device combines functionalities
of thermoelectric and spintronic devices.

As an example of some of the possible functionalities of the circuit
element, we propose a realization of magnetocaloritronic cooling and
power generation. We further study the efficiency of the magnetocaloritronic
cooling and power generation which leads us to the introduction of
the magnetocaloritronic figure of merit by analogy to the thermoelectric
figure of merit. Our estimates of the magnetocaloritronic figure of
merit for Py and MnSi give very small numbers unusable for applications.
However, we speculate that one can achieve better efficiencies using
the ferromagnetic insulators in which the heat transferred by spin
waves will better couple to the texture dynamics in the absence of
dissipation related to the electron-hole continuum.

This work was supported in part by the Alfred P. Sloan Foundation,
DARPA and NSF under Grant No. DMR-0840965.

\bibliographystyle{apsrev}
\bibliography{MagMachine}

\end{document}